\documentstyle[multicol,aps,psfig]{revtex}
\renewcommand{\narrowtext}{\begin{multicols}{2}
\global\columnwidth20.5pc\noindent}
\renewcommand{\widetext}{\end{multicols}
\global\columnwidth42.5pc}
\begin{document}
\draft
\preprint{12 September 2000}
\title{Significance of the direct relaxation process in the
       low-energy spin dynamics of a one-dimensional ferrimagnet
       NiCu(C$_7$H$_6$N$_2$O$_6$)(H$_2$O)$_3$$\cdot$2H$_2$O}
\author{Shoji Yamamoto}
\address{Department of Physics, Okayama University,
         Tsushima, Okayama 700-8530, Japan}
\date{Received 18 September 2000}
\maketitle
\begin{abstract}
In response to recent nuclear-magnetic-resonance measurements on a
ferrimagnetic chain compound
NiCu(C$_7$H$_6$N$_2$O$_6$)(H$_2$O)$_3$$\cdot$2H$_2$O
[Solid State Commun. {\bf 113} (2000) 433],
we calculate the nuclear spin-lattice relaxation rate $1/T_1$ in
terms of a modified spin-wave theory.
Emphasizing that the dominant relaxation mechanism arises from the
direct (single-magnon) process rather than the Raman (two-magnon)
one, we explain the observed temperature and applied-field
dependences of $1/T_1$.
Ferrimagnetic relaxation phenomena are generally discussed and novel
ferrimagnets with extremely slow dynamics are predicted.
\end{abstract}
\narrowtext

\section{Introduction}\label{S:I}

   Design of molecule-based ferromagnets has been one of the most
exciting subject in Materials science.
One can in principle obtain molecular ferromagnets by assembling
molecular bricks so as to construct a low-dimensional system with a
magnetic ground state and then coupling the chains or the layers
again in a ferromagnetic fashion \cite{Mill69}.
In the naivest attempt to obtain a magnetic ground state, we may
couple the nearest-neighbor magnetic centers ferromagnetically
\cite{Cair01}.
However, it is often difficult to realize the symmetry conditions
favoring the parallel alignment of local spins.
The difficulty was overcome by the introduction of a new
concept$-$antiferromagnetically coupled polymetallic systems with
irregular spin-state structures \cite{Kahn89}.
Ordered bimetallic chain compounds were thus synthesized and since
then the magnetic properties of ferrimagnetic Heisenberg chains have
extensively been investigated
\cite{Verd44,Dril13,Alca67,Pati94,Breh21,Yama24,Yama95}.

   Recently nuclear magnetic resonance for $^1$H nuclei has been
performed in a ferrimagnetic chain compound and stimulative
temperature and applied-field dependences of the proton spin-lattice
relaxation rate $1/T_1$ has been reported \cite{Fuji33}.
The authors analyzed $1/T_1$ in terms of the naivest spin-wave
theory \cite{Yama11} but could not successfully interpret the
characteristic field dependence, which looks like
$1/T_1\propto 1/\sqrt{H}$.
We here point out that {\it their argument broke down because they
attributed the dominant relaxation mechanism to the Raman
(two-magnon) process}.
Modifying the naivest spin-wave theory so as to fully describe the
thermodynamics, we demonstrate that {\it the direct (single-magnon)
relaxation process can be effective in Heisenberg ferrimagnets}. 
We further discuss potential ferrimagnetic spin dynamics arising from
the characteristic twofold excitations \cite{Yama10}.

\section{Spin-Wave Approach}\label{S:SWA}

   The measured compound
NiCu(pba)(H$_2$O)$_3$$\cdot$$2$H$_2$O
(pba $=$ $1,3$-propylenebis(oxamato) $=$ C$_7$H$_6$N$_2$O$_6$)
\cite{Pei38} consists of ordered bimetallic chains with
alternating octahedral Ni$^{2+}$ and square-pyramidal Cu$^{2+}$ ions
bridged by oxamato groups.
The one-dimensional character holds down to $7[\mbox{K}]$ under
the exchange coupling $J/k_{\rm B}\simeq 121[\mbox{K}]$.
The $g$ factors of the $S=1$ and $s=\frac{1}{2}$ spins are both close
to $2$ \cite{Hagi09}.
Thus the material is reasonably described by the one-dimensional
mixed-spin Heisenberg Hamiltonian
\begin{equation}
   {\cal H}
      =J\sum_{j=1}^N
        \left(
         \mbox{\boldmath$S$}_{j} \cdot \mbox{\boldmath$s$}_{j}
        +\mbox{\boldmath$s$}_{j} \cdot \mbox{\boldmath$S$}_{j+1}
        \right)
      -g\mu_{\rm B} H\sum_{j=1}^N(S_j^z+s_j^z)\,,
   \label{E:H}
\end{equation}
where
$\mbox{\boldmath$S$}_{j}$ and $\mbox{\boldmath$s$}_{j}$ are
respectively spin-$S(\equiv 1)$ and spin-$s(\equiv\frac{1}{2})$
operators at the $j$th elementary cell.
The spin-lattice relaxation rate is given by
\begin{eqnarray}
   \frac{1}{T_1}
   &=&\frac{4\pi(g\mu_{\rm B}\hbar\gamma_{\rm N})^2}
          {\hbar\sum_n{\rm e}^{-E_n/k_{\rm B}T}}
     \sum_{n,m}{\rm e}^{-E_n/k_{\rm B}T}
   \nonumber \\
   &\times&
     \big|
      \langle m|{\cal H}_{\rm int}|n\rangle
     \big|^2
     \,\delta(E_m-E_n-\hbar\omega_{\rm N})\,,
\label{E:T1def}
\end{eqnarray}
where $\omega_{\rm N}\equiv\gamma_{\rm N}H$ is the Larmor frequency
of the nuclei with the gyromagnetic ratio $\gamma_{\rm N}$ and the
summation $\sum_n$ is taken over all the electronic eigenstates
$|n\rangle$ with energy $E_n$.
Assuming the hyperfine interaction to be isotropic, we may represent
${\cal H}_{\rm int}$ as
\begin{equation}
   {\cal H}_{\rm int}
    =\sum_j
     \bigr(
      A\mbox{\boldmath$I$}_{j}\cdot\mbox{\boldmath$S$}_{j}
     +B\mbox{\boldmath$I$}_{j}\cdot\mbox{\boldmath$s$}_{j}
     \bigl)\,,
\end{equation}
with the nuclear spin operators $\mbox{\boldmath$I$}_{j}$ and the
dipolar coupling constants $A$ and $B$.

   In order to calculate $1/T_1$ in practice, we introduce the
bosonic operators for the spin deviation as
\begin{equation}
   \left.
   \begin{array}{lll}
      S_j^+=(2S-a_j^\dagger a_j)^{1/2}a_j\,,&
      S_j^z=S-a_j^\dagger a_j\,,\\
      s_j^+=b_j^\dagger(2s-b_j^\dagger b_j)^{1/2}\,,&
      s_j^z=-s+b_j^\dagger b_j\,.
   \end{array}
   \right.
   \label{E:HPT}
\end{equation}
Then, assuming $O(S)=O(s)$, we can expand ${\cal H}_{\rm int}$ with
respect to $1/S$ and obtain $1/T_1=\sum_{l=1,2,\cdots} 1/T_1^{(l)}$
\cite{Beem59}, where $1/T_1^{(l)}$ is the $l$-magnon relaxation rate
within the first-order spin-lattice relaxation process and is given
by replacing ${\cal H}_{\rm int}$ by its $O(S^{1-l/2})$-component in
Eq. (\ref{E:T1def}).
On the other hand, within the up-to-$O(S^0)$ approximation, the
electronic Hamiltonian (\ref{E:H}) can be diagonalized as
\cite{Yama33}
\begin{equation}
   {\cal H}\simeq{\cal H}_{\rm SW}
     =E_{\rm g}+\sum_k
     \left(
      {\widetilde\omega}_k^- \alpha_k^\dagger \alpha_k
     +{\widetilde\omega}_k^+ \beta_k^\dagger  \beta_k
     \right)\,.
   \label{E:HSW}
\end{equation}
Here, $E_{\rm g}$ is the spin-wave ground-state energy, whereas
$\alpha_k^\dagger$ and $\beta_k^\dagger$ create the spin waves of
ferromagnetic and antiferromagnetic aspects \cite{YamaEPJB},
respectively, and are related with the sublattice bosons via
$\alpha_k^\dagger
 =a_k^\dagger{\rm cosh}\theta_k+b_k{\rm sinh}\theta_k$ and
$\beta_k^\dagger
 =a_k{\rm sinh}\theta_k+b_k^\dagger{\rm cosh}\theta_k$ with
$a_k^\dagger
 =(1/\sqrt{N})\sum_j{\rm e}^{-{\rm i}k(j-1/4)}a_j^\dagger$,
$b_k^\dagger
 =(1/\sqrt{N})\sum_j{\rm e}^{ {\rm i}k(j+1/4)}b_j^\dagger$, and
${\rm tanh}(2\theta_k)=2\sqrt{Ss}{\rm cos}(k/2)/(S+s)$.
We here take twice the lattice constant as unity.
The dispersion relations are given by
\begin{equation}
   {\widetilde\omega}_k^\pm=\omega_k^\pm-\delta\omega_k^\pm\,,
\end{equation}
where
\begin{eqnarray}
   \omega_{k}^\pm
   &=&
    \omega_k\pm(S-s)J\mp g\mu_{\rm B}H\,,
   \\
    \delta\omega_k^\pm
   &=&
    2(S+s){\mit\Gamma}_1
    \frac{\sin^2(k/2)}{\omega_k}
   -\frac{{\mit\Gamma}_2}{\sqrt{Ss}}
    [\omega_k\pm(S-s)]\,,
\end{eqnarray}
with
\begin{eqnarray}
   \omega_k
   &=&
    J\sqrt{(S-s)^2+4Ss\sin^2(k/2)}\,,
   \\
   {\mit\Gamma}_1
   &=&
    \frac{1}{2N}\sum_k
     \left(
      {\displaystyle\frac{S+s}{\omega_k}-1}
     \right)\,,
   \\
   {\mit\Gamma}_2
   &=&
    \frac{1}{N}\sum_k
    \frac{\sqrt{Ss}\,{\rm cos}^2(k/2)}{\omega_k}\,.
\end{eqnarray}
\vskip 2mm
\begin{figure}
\begin{flushleft}
\ \mbox{\psfig{figure=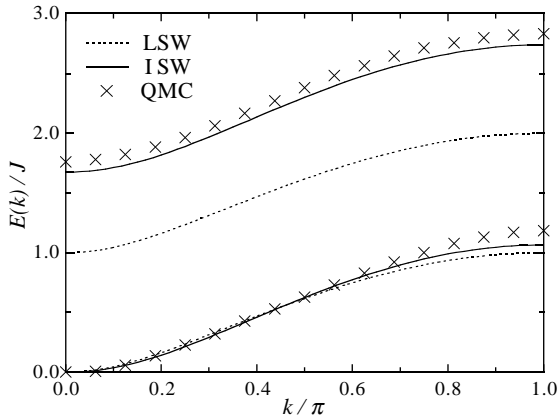,width=86mm,angle=0}}
\end{flushleft}
\vskip 0mm
\caption{Dispersion relations of the linear- (dotted lines; LSW) and
         interacting- (solid lines; ISW) spin-wave excitation
         energies compared with the lowest-energy bands in the
         subspaces of $\sum_{i=1}^N(S_i^z+s_i^z)=(S-s)N\mp 1$
         estimated by a quantum Monte Carlo method ($\times$; QMC)
         for $(S,s)=(1,\frac{1}{2})$.}
\label{F:Ek}
\end{figure}
\noindent
The linear spin waves are characterized by $\omega_k^\pm$, while the
interacting spin waves by ${\widetilde\omega}_k^\pm$ with the
$O(S^0)$ quantum corrections $\delta\omega_k^\pm$.
We plot $\omega_k^\pm$ and ${\widetilde\omega}_k^\pm$ in Fig.
\ref{F:Ek}.
The spin waves describe the elementary excitations satisfactorily,
especially the low-energy excitations near the zone center, which
must be the most relevant to the nuclear spin relaxation.
The antiferromagnetic excitation energies are underestimated by the
linear spin waves but are significantly corrected in consideration of
the interactions between them.
All the following calculations are based on the interacting-spin-wave
dispersions ${\widetilde\omega}_k^\pm$.

   The $\delta$ function in Eq. (\ref{E:T1def}) insures the
conservation of energy in the transition.
Considering the significant difference between the electronic and
nuclear energy scales ($\hbar\omega_{\rm N}\alt 10^{-5}J$), the
relevant spin-wave excitations are strongly limited.
Only the small-momentum ferromagnetic spin waves contribute to the
direct process.
In the Raman process, both ferromagnetic and antiferromagnetic spin
waves are effective but interband transitions are still irrelevant.
Scatterings between ferromagnetic and antiferromagnetic spin waves
are relevant to the higher-order processes containing three or more
magnons.
We show a few leading terms of the relaxation rate:
\begin{eqnarray}
   \frac{1}{T_1^{(1)}}
   &\simeq&
     \frac{4(g\mu_{\rm B}\hbar\gamma_{\rm N})^2}{NJ\hbar}
     \sqrt{\frac{2(S-s)}{Ss\hbar\omega_{\rm N}/J}}
   \nonumber \\
   &\times&
    (\sqrt{S}A{\rm cosh}\theta_0-\sqrt{s}B{\rm sinh}\theta_0)^2
    (n_0^-+1)\,,
   \label{E:T1-1} \\
   \frac{1}{T_1^{(2)}}
   &\simeq&
     \sum_k
     \frac{4(g\mu_{\rm B}\hbar\gamma_{\rm N})^2(S-s)}
          {NJ\hbar\sqrt{(Ssk)^2+2(S-s)Ss\hbar\omega_{\rm N}/J}}
   \nonumber \\
   &\times&
   \bigl[
    (A{\rm cosh}^2\theta_k-B{\rm sinh}^2\theta_k)^2
     n_k^-(n_k^- +1)
   \nonumber \\
   &&+
    (A{\rm sinh}^2\theta_k-B{\rm cosh}^2\theta_k)^2
     n_k^+(n_k^+ +1)
   \bigr]\,,
   \label{E:T1-2} \\
   \frac{1}{T_1^{(3)}}
   &\simeq&
     \sum_{k,k'}
     \frac{(g\mu_{\rm B}\hbar\gamma_{\rm N})^2(S-s)}
          {2N^2 J\hbar
           \sqrt{S^2s^2(k^2+k'^2)+2(S-s)Ss\hbar\omega_{\rm N}/J}}
   \nonumber \\
   &\times&
   \Bigl[
     \bigl(
     \frac{A}{\sqrt{S}}
     {\rm cosh}\theta_{k}
     {\rm cosh}\theta_{k'}
     {\rm cosh}\theta_{k+k'}
    - \frac{B}{\sqrt{s}}
     {\rm sinh}\theta_{k}
   \nonumber \\
   &&\times
     {\rm sinh}\theta_{k'}
     {\rm sinh}\theta_{k+k'}
     \big)^2
     n_{k}^-n_{k'}^-(n_{k+k'}^-+1)
   \nonumber \\
   &+&
    4\bigl(
     \frac{A}{\sqrt{S}}
     {\rm cosh}\theta_{k}
     {\rm sinh}\theta_{k'}
     {\rm sinh}\theta_{k+k'}
    - \frac{B}{\sqrt{s}}
     {\rm sinh}\theta_{k}
   \nonumber \\
   &&\times
     {\rm cosh}\theta_{k'}
     {\rm cosh}\theta_{k+k'}
     \big)^2
     n_{k}^-n_{k'}^+(n_{k+k'}^++1)
   \Bigr]\,,
   \label{E:T1-3}
\end{eqnarray}
where
$n_k^-\equiv\langle\alpha_k^\dagger\alpha_k\rangle$ and
$n_k^+\equiv\langle \beta_k^\dagger\beta_k \rangle$ are the
distribution functions of the spin waves.
In general $1/T_1^{(l)}$ rapidly decreases as a function of $l$.
However, a slight anisotropy, for example, turns off the direct
process and therefore makes the Raman process effective.
In such a case, the three-magnon-process relaxation rate can also be
effecitve, being enhanced via the exchange interaction, at high
temperatures \cite{Beem59,Pinc98}.

\begin{figure}
\begin{flushleft}
\quad\mbox{\psfig{figure=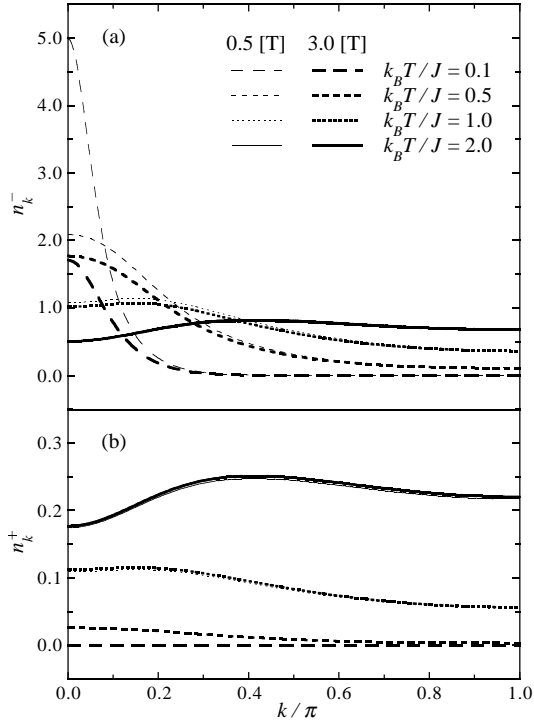,width=82mm,angle=0}}
\end{flushleft}
\vskip 0mm
\caption{The momentum distribution functions of the ferromagnetic (a)
         and antiferromagnetic (b) spin waves at various values of
         temperature and the applied magnetic field in the case of
         $(S,s)=(1,\frac{1}{2})$.}
\label{F:nk}
\end{figure}

   Now the problem is reduced to the evaluation of $n_k^\pm$.
The naivest thermodynamics defining the partition function as
$Z={\rm Tr}[{\rm e}^{-{\cal H}_{\rm SW}/k_{\rm B}T}]$ ends in the
divergence of $n_k^\pm$ with increasing temperature.
Thus we here employ the modified spin wave theory for ferrimagnets
\cite{Yama33,Yama08} and obtain the spin-wave distribution functions
as
\begin{equation}
   n_k^\pm
    =\frac{1}
     {{\rm e}^{[{\widetilde\omega}_k^\pm-\mu(S+s)J/\omega_k]
               /k_{\rm B}T}-1}\,,
\end{equation}
where $\mu$ is a Lagrange multiplier controlling the staggered
magnetization and is self-consistently determined under the
constraint $\sum_k\sum_{\sigma=\pm}n_k^\sigma/\omega_k=N$.
We show in Fig. \ref{F:nk} the thus-obtained $n_k^\pm$ as functions
of $k$.
At low temperatures, $n_k^-$ exhibits a pronounced peak at $k=0$,
whereas $n_k^+$ stays negligibly small in comparison with $n_k^-$,
because the ferromagnetic and antiferromagnetic excitations are
respectively gapless and gapped from the ground state.
Even at high temperatures, $n_k^+$ remains less dominant due to the
large gap
${\widetilde\omega}_{k=0}^+=2J(S-s)(1+{\mit\Gamma}_2/\sqrt{Ss})$ with
${\mit\Gamma}_2=0.4777$ for $(S,s)=(1,\frac{1}{2})$.
The applied field, slightly shifting the spin-wave excitation
energies, reduces $n_k^-$ and enhances $n_k^+$.
\vskip 2mm
\vskip 0mm

\begin{figure}
\begin{flushleft}
\ \mbox{\psfig{figure=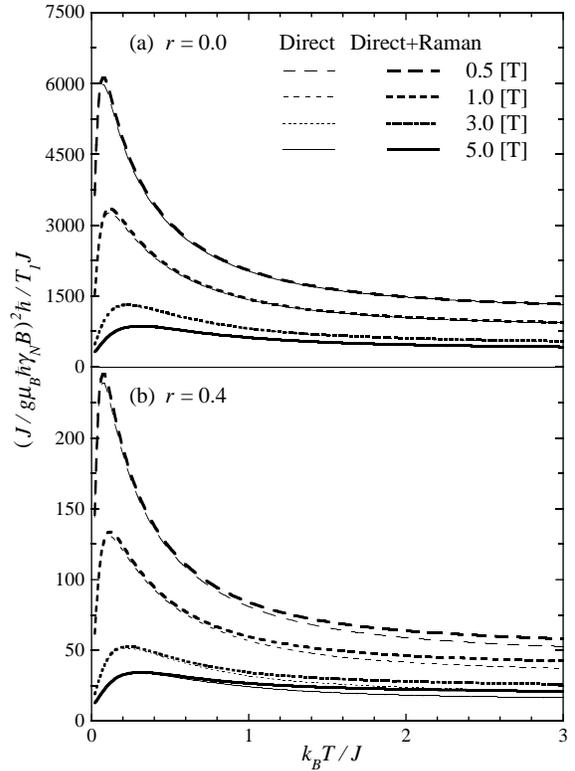,width=88mm,angle=0}}
\end{flushleft}
\vskip 0mm
\caption{Temperature dependences of the direct-process and
         Raman-process relaxation rates at various values of the
         applied magnetic field in the case of
         $(S,s)=(1,\frac{1}{2})$.
         $1/T_1^{(1)}$ is plotted by thin lines, while
         $1/T_1^{(1)}+1/T_1^{(2)}$ by thick lines.}
\label{F:T1TDR}
\end{figure}

   Let us observe general behavior of $1/T_1$ in Fig. \ref{F:T1TDR}.
We have learned in Fig. \ref{F:nk} that the ferromagnetic and
antiferromagnetic spin waves respectively contribute decreasing and
increasing components as functions of temperature to the relaxation
rate.
Therefore, $1/T_1^{(1)}$ results in a monotonically decreasing
function of $T$, whereas $1/T_1^{(2)}$ contains increasing components
as well as overwhelming decreasing ones.
Here is a fascinating parameter to be adjusted, $r\equiv A/B$.
Dipolar coupling constants are proportional to the inverse cube of
the distance between the interacting nuclear and electronic spins
and therefore $r$ is quite sensitive to the crystalline structure.
Considering the predominance of the small-momentum excitations, we
find there is a special point of $r=s/S$, where the dominant
relaxation mechanism mediated by the ferromagnetic spin waves does
not work at all in every $l$-magnon process and thus the Raman
process mediated by the antiferromagnetic spin waves becomes the
leading relaxation mechanism.
That is why as $r$ approaches $s/S$, the Raman contribution increases
emphasizing the increasing behavior as a function of $T$.

\section{Explanation of the Experiment}\label{S:EE}

   We apply the theory to the proton spin relaxation \cite{Fuji33} in
NiCu(pba)(H$_2$O)$_3$$\cdot$$2$H$_2$O.
Since the protons mainly contributing to $1/T_1$ turn out to lie
in the pba groups near Cu ions, we hear set $r$ smaller than unity.

\begin{figure}
\begin{flushleft}
\ \mbox{\psfig{figure=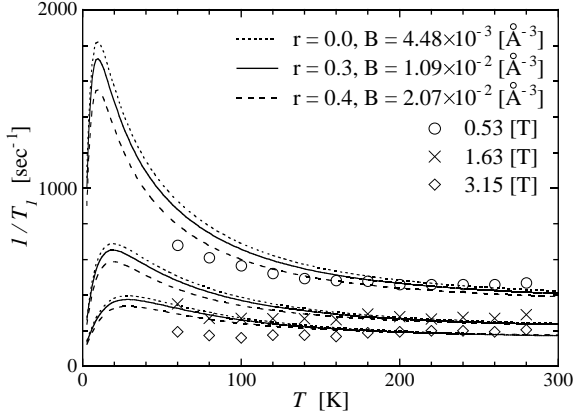,width=88mm,angle=0}}
\end{flushleft}
\vskip 0mm
\caption{Temperature dependences of the direct-process relaxation
         rate plus the Raman-process one, $1/T_1^{(1)}+1/T_1^{(2)}$,
         at various values of the applied magnetic field and the
         structural parameters in the case of
         $(S,s)=(1,\frac{1}{2})$, assuming that
         $J/k_{\rm B}=121[\mbox{K}]$.
         The calculations are shown by lines, where
         $H=0.53, 1.63, 3.15[\mbox{T}]$ from the top to the bottom,
         while the measurements by symbols.}
\label{F:T1Texp}
\end{figure}
\vskip -2mm
\begin{figure}
\begin{flushleft}
\ \mbox{\psfig{figure=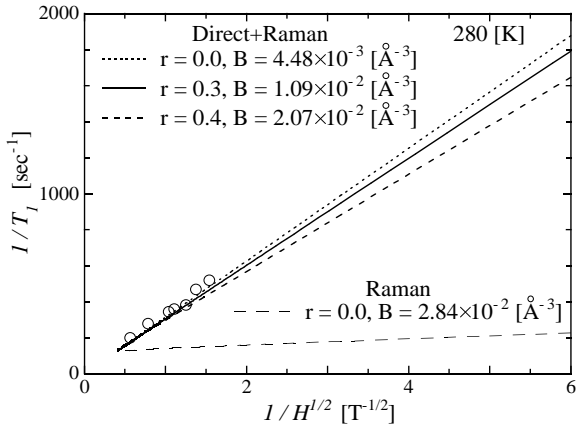,width=90mm,angle=0}}
\end{flushleft}
\vskip 0mm
\caption{Field dependences of the direct-process relaxation
         rate plus the Raman-process one, $1/T_1^{(1)}+1/T_1^{(2)}$,
         at various values of the structural parameters in the case
         of $(S,s)=(1,\frac{1}{2})$, assuming that
         $J/k_{\rm B}=121[\mbox{K}]$.
         Temperature is set equal to $280[\mbox{K}]$.
         The calculations are shown by lines, while the measurements
         by $\circ$.
         A calculation for  $1/T_1^{(2)}$ is also shown for
         reference.}
\label{F:T1Hexp}
\end{figure}
\vskip 0mm

   We analyze the observed temperature dependences of $1/T_1$
\cite{Fuji33} in Fig. \ref{F:T1Texp}.
Though no parametrization ends in a total agreement with the
measurements, the calculation can reproduce them fairly well.
In the present sample, the relevant protons have a rather wide
distribution around Cu ions and therefore the distances between the
interacting proton and electron spins, which we refer to as $a$ and
$b$ for $S=1$ and $s=\frac{1}{2}$, respectively, can not be
determined definitely.
If we assume that $A\sim 1/a^3$ and $B\sim 1/b^3$,
$4.48\times 10^{-3}[\mbox{\AA}^{-3}]$ for $B$ may be too small, while
$0.4$ for $r$ be too large.
As the applied field increases, the recovery curve of the spin-echo
decay more and more deviates from a single-exponential function
\cite{Fuji}.
The experimental estimates of $1/T_1$ at $H\agt 1[\mbox{T}]$ seem to
contain larger uncertainty.

   Field dependence of $1/T_1$ was measured in more detail at
$280[\mbox{K}]$ \cite{Fuji33} and is analyzed in Fig. \ref{F:T1Hexp}.
We again find that the calculations with small $r$ can generally be
fitted to the measurements.
We emphasize that the linear dependence of $1/T_1$ on $1/\sqrt{H}$
with a steep slope can never be attributed to the Raman process but
be described in consideration of the direct process.
As the most significant field dependence is obtained at $r=0$ in each
relaxation process, the Raman contribution to the field dependence
turns out much smaller than that from the direct process.
The three-magnon contribution, even if it is enhanced via the
exchange interaction, is still less relevant.
In general, the momentum integral in the calculation of $1/T_1$
weakens the linear field dependence and even leads to logarithmic one
at high fields.

   We should be reminded that the present spin-wave theory, which is
free from both quantum and thermal divergences of the sublattice
magnetizations, is highly successful but its ability to describe
thermodynamics is still restricted quantitatively \cite{YamaEPJB}.
Considering minor corrections to the theory as well, arising from
distribution of proton spins, slight difference between $g$ factors,
weak interchain interactions and so forth, we are allowed to conclude
that the direct process plays a leading role in the nuclear spin
relaxation of NiCu(pba)(H$_2$O)$_3$$\cdot$$2$H$_2$O.

\section{Concluding Remarks}\label{S:CC}

   The dominant relaxation mechanism for the title compound has been
shown to arise from the direct process.
It is at the same time the demonstration that the material is a
really isotropic magnet.

   Another harvest we have obtained is the prediction of novel
ferrimagnets with extremely slow dynamics.
Under the crystalline structure with $r=s/S$, the fast relaxation
processes mediated purely by the ferromagnetic spin waves hardly work
and thus the nuclear spin is very slow in relaxing especially at low
temperatures.

\acknowledgments

   I thank Dr. N. Fujiwara for directing my attention to the
present topic and helpful comments.
I am grateful to Prof. T. Goto as well for useful discussion.
This work was supported by the Japanese Ministry of Education,
Science, and Culture.
The numerical computation was done using the facility of the
Supercomputer Center, Institute for Solid State Physics, University
of Tokyo.

\widetext
\end{document}